\documentclass[manuscript]{acmart}
\usepackage{booktabs}
\usepackage{tabularx}
\usepackage{pifont}
\newcommand{\supported}{\ding{51}}
\setcopyright{none}
\renewcommand\footnotetextcopyrightpermission[1]{}
\acmConference[IUI '27 Companion]{Companion Proceedings of the 32nd International Conference on Intelligent User Interfaces}{February 8--11, 2027}{Helsinki, Finland}
\acmYear{2027}
\copyrightyear{2027}
\acmDOI{}
\acmISBN{}
\title{GAZEAIDE: Recording Gaze Context for Research on AI-Assisted Development}
\author{Heidi Albarazi}
\affiliation{\institution{JetBrains Research}\city{Munich}\country{Germany}}\email{heidi.albarazi@jetbrains.com}
\author{Ilya Zakharov}\authornote{Corresponding author.}\affiliation{\institution{JetBrains Research}\city{Belgrade}\country{Serbia}}\email{ilia.zaharov@jetbrains.com}
\author{Ekaterina Koshchenko}
\affiliation{\institution{JetBrains Research}\city{Amsterdam}\country{Netherlands}}\email{ekaterina.koshchenko@jetbrains.com}
\author{Agnia Sergeyuk}
\affiliation{\institution{JetBrains Research}\city{Belgrade}\country{Serbia}}\email{agnia.sergeyuk@jetbrains.com}

\begin{abstract}
Developers increasingly review code they did not write, much of it generated by AI. Where they look while reading may indicate what is relevant to them, but tools rarely have access to this information. We demonstrate GAZEAIDE, an open-source research prototype for collecting it. Using a webcam, GAZEAIDE estimates which parts of the code a developer looks at and for how long, and records this viewing history together with the text shown on screen. Such context can serve different purposes: providing visual feedback on what has been viewed, guiding code review, and preparing context for a question, which a developer can choose to send to an AI assistant. An IntelliJ plugin records work in the editor; a browser companion records assistant conversations. Attendees can record a short session, replay their viewing history, and try these scenarios.
\end{abstract}


\begin{document}
\maketitle
\raggedbottom

\section{Motivation: use the context produced during reading}
AI-assisted development involves reading and checking code that developers did not fully author. Asking for help requires developers to recognize a need, formulate a question, and convey the relevant material. The motivation for GAZEAIDE, short for \emph{Gaze-Augmented Interaction for Developer Environments}, is to explore an interaction layer that can use context already produced during reading, potentially reducing the additional effort needed to communicate what is relevant.

Consider a developer investigating why a name containing only spaces is accepted instead of rejected. They read the caller that trims spaces, the validator that accepts an empty value, and a check that expects a default guest label. These passages may help an assistant explain the behavior. Recording which versions were viewed would let a researcher trace why each passage entered the assistant's context.

\begin{figure}[t]
\centering
\includegraphics[width=0.8\linewidth]{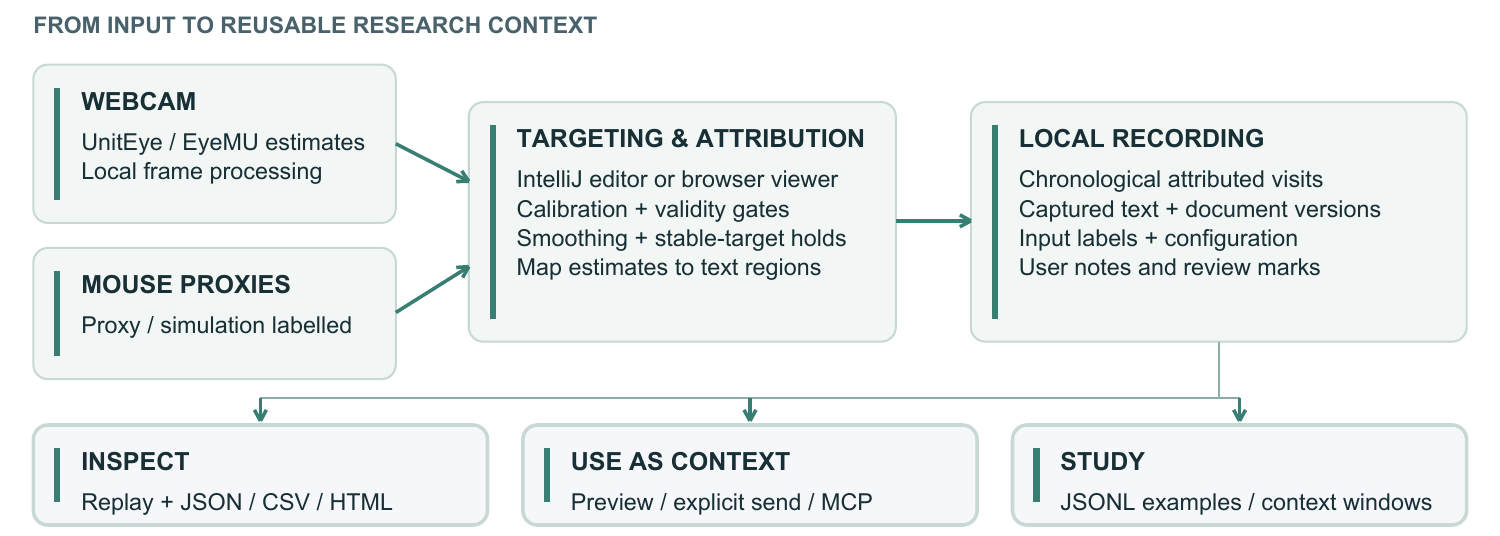}
\caption{GAZEAIDE separates sensing and attribution, local recording, and downstream uses. Recordings preserve attributed visits and displayed text. Assistant requests and remote upload require separate explicit actions.}
\Description{Webcam estimates and labelled mouse proxies feed targeting and attribution in an IntelliJ editor or browser viewer. Calibration, validity gates, smoothing, and stable-target holds map estimates to text regions. Local recording preserves chronological visits, text, versions, input labels, notes, and review marks. The recording supports replay and JSON, CSV, or HTML inspection; context preview, explicit submission, or MCP retrieval; and JSONL examples or context windows for research.}
\label{fig:workflow}
\end{figure}

GAZEAIDE records visits, reconstructs encountered content, and exports context linked to assistant requests (Figure~\ref{fig:workflow}). Because treating every look as a command risks unintended activation~\cite{jacob1991eye}, viewing supplies context while users control question submission.

\section{Positioning among gaze-aware tools}
\paragraph{Gaze-aware developer tools}
Gaze has served both as an observation channel and as an input to developer tools. iTrace passively maps eye movements to software artifacts within the IDE~\cite{shaffer2015itrace}; iTrace-Atom and gazel extend capture to source edits and reconstruct gaze-linked file versions~\cite{fakhoury2021gazel}. EyeDE uses intentional looks at in-code triggers to open documentation or navigate to method declarations~\cite{glucker2014eyede}. Cheng et al. share reviewers' gaze through live line highlighting and visual links between inspected passages~\cite{cheng2022collaborative}.

CodeGRITS coordinates gaze, IDE events, and screen recording in JetBrains IDEs, with activity labels, structured export, and a real-time data API~\cite{tang2024codegrits}. Its mapping from gaze coordinates to code locations and syntax elements is an antecedent for GAZEAIDE's editor attribution.

\paragraph{Gaze-informed AI assistance}
GazeCopilot converts gaze metrics into prompt text for AI-supported code comprehension and readability; a keyboard shortcut initiates assistance~\cite{elfares2025gazecopilot}. GazeGPT uses gaze to identify the physical object addressed by a spoken query through smart eyewear~\cite{konrad2024gazegpt}. EmBARDiment maintains gaze-selected text context across XR windows and combines it with an explicit verbal request~\cite{bovo2025embardiment}. These systems show how gaze can enrich AI input while users initiate requests. They represent different forms of context: gaze metrics over code, objects in a scene, and previously encountered text.

Among the systems compared in Table~\ref{tab:comparison}, GAZEAIDE combines gaze-to-code mapping, gaze-driven visual feedback, gaze-informed assistant input, and machine-readable export in one workflow. It links captured text and viewing history to each assistant request it prepares.

\begin{table}[t]
\caption{Documented roles and capabilities (\supported{} supported; -- not reported or outside scope). Gaze-driven User Interface (UI) covers navigation/highlighting. Export means machine-readable gaze/context records; GazeCopilot uses CodeGRITS. Gaze in AI input does not imply automatic requests. GAZEAIDE's row spans both editions.}
\label{tab:comparison}
\centering
\small
\setlength{\tabcolsep}{3pt}
\begin{tabularx}{\linewidth}{@{}p{0.24\linewidth}X*{4}{>{\centering\arraybackslash}p{0.105\linewidth}}@{}}
\toprule
System & Main role & \shortstack{Gaze to\\code} & \shortstack{Gaze-driven\\UI} & \shortstack{Gaze in\\AI input} & Export \\
\midrule
iTrace / gazel~\cite{shaffer2015itrace,fakhoury2021gazel} & Capture; edited source & \supported & -- & -- & \supported \\
EyeDE~\cite{glucker2014eyede} & Code navigation & \supported & \supported & -- & -- \\
Cheng et al.~\cite{cheng2022collaborative} & Shared review attention & \supported & \supported & -- & -- \\
CodeGRITS~\cite{tang2024codegrits} & IDE and gaze collection & \supported & -- & -- & \supported \\
GazeCopilot~\cite{elfares2025gazecopilot} & Gaze-metric prompting & \supported & -- & \supported & \supported \\
GazeGPT~\cite{konrad2024gazegpt} & Queries about objects & -- & -- & \supported & -- \\
EmBARDiment~\cite{bovo2025embardiment} & XR text-context memory & -- & -- & \supported & -- \\
\textbf{GAZEAIDE} & Replayable gaze context & \supported & \supported & \supported & \supported \\
\bottomrule
\end{tabularx}
\end{table}

\section{From viewing to reusable context}
\paragraph{Two editions for a broader interaction setting}
GAZEAIDE began as an IntelliJ plugin connecting gaze estimates to editor lines and code regions. The research scope then expanded to assistant explanations, generated code, and earlier conversation turns, which also shape subsequent questions but fall outside editor capture. We therefore added a standalone browser companion. Its viewer displays source files, text, and locally available Codex history, preserving conversation provenance and historical diffs. It also imports plugin recordings for inspection.

\paragraph{Record and reconstruct what was encountered}
Recording starts with an input mode, scenario, and optional participant/task/condition codes. Webcam estimation uses UnitEye~\cite{wagner2024uniteye} and EyeMU~\cite{kong2021eyemu}, with local processing in a browser or the Java Chromium Embedded Framework (JCEF) processing and calibration. Nine calibration targets fit a user-specific correction, followed by five held-out targets that report spatial error and jitter, including vertical error relative to editor line height. Smoothing and stable-target holds govern attribution, while region-based assistant context remains inspectable before submission. These features support setup assessment and exploratory context collection; line-level attribution during natural reading remains to be evaluated. Mouse proxy and simulation retain distinct input labels.

The system automatically records \emph{line visits}: intervals attributed to a displayed line, with timing, text, input source, and available region snapshots and document versions. Pauses and invalid input interrupt attribution. Visits estimate exposure, not fixations or understanding. Task recordings contain neither camera video nor a continuous raw gaze stream.

\begin{figure}[t]
\centering
\includegraphics[width=0.8\linewidth]{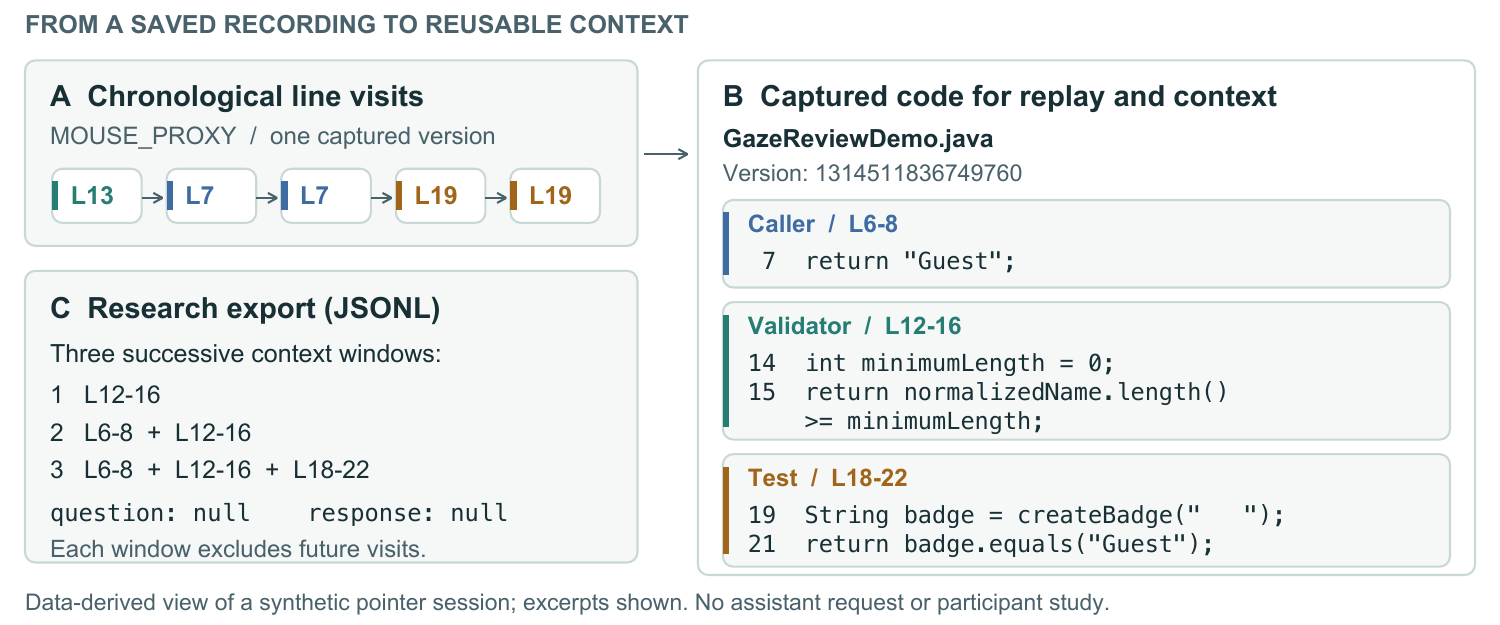}
\caption{A synthetic \texttt{MOUSE\_PROXY} companion recording links (A) chronological visits to (B) saved code and (C) three successive context windows. Derived from actual recording/export data; null questions and responses indicate viewing only. No participant results or accuracy measurement.}
\Description{Five chronological mouse-proxy visits to lines 13, 7, 7, 19, and 19 link to saved excerpts from GazeReviewDemo.java. The caller returns Guest, the validator accepts a name of length zero, and a test checks a spaces-only name against Guest. Three successive JSONL context windows contain the validator, then the caller and validator, then all three regions. Each window excludes future visits and retains null question and response fields. All excerpts share one captured document version.}
\label{fig:recording}
\end{figure}

Researchers can replay saved text after files change, inspect passages, and add notes or explicit review marks. Figure~\ref{fig:recording} links visits to a validator, caller, and test: captured regions retain their whitespace-handling code, while successive exports reconstruct the context available at each window’s end. Exports include structured JSON, CSV for analysis, and self-contained HTML replay outside the Integrated Development Environment (IDE).

\paragraph{Inspect assistant context and export research records}
GAZEAIDE prepares a bounded context packet from the selected target, visible content, and recently visited regions. The selected target and visible viewport take priority; other regions require at least 200 ms of accumulated exposure and are ordered by their most recent visit within a ten-minute window, considering at most 5,000 visits. The packet retains at most eight regions, with excerpts capped at 1,800 characters and a 12,000-character budget for the serialized context. It preserves excerpts, source/version and line identifiers, input provenance, recency, and exposure, and reports truncation. Visible-but-unvisited text is marked separately. Selection uses no future visits and remains an unvalidated model of relevance.

Users do not attach each passage manually: context accompanies an editable question. In the plugin, \emph{Copy prompt \& open AI Chat} copies the complete prompt for the user to paste and send in native chat; delivery and replies are unobserved. The companion's send action attaches fresh context when enabled and observes replies through its own Codex conversation. Sustained viewing, or \emph{dwell}, can prepare an optional local question without calling a model. A read-only Model Context Protocol (MCP) interface lets configured clients retrieve context.

Research records link the question, exact prepared request, context, attachment state, outcome, and optional feedback. Unobserved replies remain missing; pasted answers are labelled as user transcriptions. JSONL export also supports viewing-only \emph{context windows} with null question/answer fields. These records support future comparisons of question-only, conventional, and viewing-based context.

\section{Scenarios as examples and experimental conditions}
Ten scenarios illustrate the workflow. \emph{Record viewing only} provides a baseline with visual effects and AI preparation off. Four visual-feedback examples make accumulated input visible: \emph{Follow my reading} keeps the viewed block clear while dimming its surroundings; \emph{Highlight viewed lines} tints lines after exposure; \emph{Fade previously viewed code} fades visited blocks and restores them on return; and \emph{Highlight matching names} emphasizes matching identifier text.

Two review scenarios connect history to explicit work. \emph{Gaze-assisted review} prepares explain, compare, or review questions from captured regions. \emph{Gaze-guided review walkthrough} turns captured regions into inspectable steps with notes and explicit review marks. The remaining three scenarios support configuration and analysis. \emph{Personalization} fits a smoothing/response preset through a short guided task. \emph{Personal baseline \& blink activity} optionally records webcam closure events and sampling quality across sessions; these observations do not establish fatigue. Finally, \emph{Custom features} combines effects and delays. These scenarios illustrate possible applications and study conditions, rather than demonstrated improvements in reading or review performance.

\section{Interactive demonstration and current status}
Visitors use the IntelliJ plugin to record the validation example, replay captured text, and inspect how the caller and validator enter a prepared question. They then export the recording and examine its research fields. A brief companion demonstration extends capture to assistant explanations and conversation history.

The setup uses a webcam-equipped laptop, IntelliJ/JCEF, and a local Node.js/browser companion. Recording, replay, and export require no AI account; sending requires provider access. Labelled pointer input and saved recordings provide fallbacks. The video uses mouse-proxy input and synthetic content. The demonstrated source snapshot includes setup instructions.\footnote{\href{https://github.com/JetBrains-Research/GAZEAIDE/tree/a2c6121c924b88f23d14148cb00209fd1fce1b1b}{GAZEAIDE source snapshot, commit a2c6121}.} Recordings are local by default and may contain sensitive code, conversations, and behavioral data. Sending attached context discloses it to the provider; configured MCP clients can retrieve published context. Remote study collection defaults off.

\paragraph{Use of generative AI}
OpenAI Codex assisted with implementation, software checks, manuscript revision, figure code, and video assembly; narration is synthetic. The authors take responsibility for the system and submission.

\clearpage
\bibliographystyle{ACM-Reference-Format}
\bibliography{references}
\end{document}